\documentclass[aps,prstab,reprint,superscriptaddress,longbibliography]{revtex4-1}
\usepackage{graphicx}
\usepackage{booktabs}
\usepackage{amssymb}
\usepackage{graphics}
\usepackage{amsmath}
\usepackage{placeins}
\usepackage{hyperref}
\usepackage{verbatim}
\usepackage[dvipsnames]{xcolor}

\begin{document}
	
	
	\title{Rapid thermal emittance and quantum efficiency mapping of a cesium telluride cathode in an rf photoinjector using multiple laser beamlets}
	
	
	\author{Lianmin Zheng}
	\affiliation{Department of Engineering Physics, Tsinghua University Beijing, Beijing 100084, People's Republic of China}
	\affiliation{High Energy Physics Division, Argonne National Laboratory, Lemont, Illinois 60439, USA}
	\author{Jiahang Shao}
	\author{Eric E. Wisniewski}
	\author{John G. Power}
	\affiliation{High Energy Physics Division, Argonne National Laboratory, Lemont, Illinois 60439, USA}
	\author{Yingchao Du}
	\email[]{dych@mail.tsinghua.edu.cn}
	\affiliation{Department of Engineering Physics, Tsinghua University Beijing, Beijing 100084, People's Republic of China}
	\author{Wanming Liu}
	\author{Charles E. Whiteford}
	\author{Manoel Conde}
	\author{Scott Doran}
	\affiliation{High Energy Physics Division, Argonne National Laboratory, Lemont, Illinois 60439, USA}
	\author{Chunguang Jing}
	\affiliation{High Energy Physics Division, Argonne National Laboratory, Lemont, Illinois 60439, USA}
	\affiliation{Euclid Techlabs LLC, Bolingbrook, Illinois 60440, USA}
	\author{Chuanxiang Tang}
	\affiliation{Department of Engineering Physics, Tsinghua University Beijing, Beijing 100084, People's Republic of China}
	

	
	\date{\today}
	
	\begin{abstract}
		Thermal emittance and quantum efficiency (QE) are key figures of merit of photocathodes, and their uniformity is critical to high-performance photoinjectors. Several QE mapping technologies have been successfully developed; however, there is still a dearth of information on thermal emittance maps. This is because of the extremely time-consuming procedure to gather measurements by scanning a small beam across the cathode with fine steps. To simplify the mapping procedure, and to reduce the time required to take measurements, we propose a new method that requires only a single scan of the solenoid current to simultaneously obtain thermal emittance and QE distribution by using a pattern beam with multiple beamlets. In this paper, its feasibility has been confirmed by both beam dynamics simulation and theoretical analysis. The method has been successfully demonstrated in a proof-of-principle experiment using an L-band radiofrequency photoinjector with a cesium telluride cathode. In the experiment, seven beamlets were generated from a microlens array system and their corresponding thermal emittance and QE varied from 0.93 to 1.14~$\mu$m/mm and from 4.6 to 8.7\%, respectively. We also discuss the limitations and future improvements of the method in this paper.
	\end{abstract}
	
	\pacs{}
	
	\maketitle
	
	\section{INTRODUCTION}\label{intro}
	
	Beam brightness, defined by current over emittance, is one of the most important properties of photoinjectors. The continuous improvement of beam brightness over the last few decades has enabled many photoinjector-based machines for scientific research, such as X-ray free electron lasers~\cite{emma2010first,ackermann2007operation}, ultrafast electron diffraction and microscopy~\cite{weathersby2015mega,li2009experimental}, Thomson scattering X-ray sources~\cite{du2013generation,gibson2010design}, and wakefield acceleration~\cite{EnglandRMP2014,XueyingPRL2019}. The beam brightness of a photoinjector heavily depends on the photocathode, because its thermal emittance (a.k.a. intrinsic emittance, denoted as $\varepsilon_\text{therm}$ or $\varepsilon_{\text{therm},n} \equiv \varepsilon_\text{therm}/\sigma _\text{laser}$ when normalized by the rms laser spot size) sets the lower boundary of beam emittance and its quantum efficiency (QE) determines the current with certain incident laser. Recently, intense studies have focused on advanced cathode research and development (R\&D) to obtain low thermal emittance and high QE~\cite{dowell2010cathode,cultrera2015cold,feng2017near,cultrera2016ultra,lee2016intrinsic,gaowei2019codeposition,Maxson2015,wang2018overcoming,musumeci2018advances}.
	
	Most cathode studies work on the average thermal emittance and QE of large areas~\cite{qian2012experimental,lee2015review,Divall2015,FilippettoCesium,Eduard2015Measurements,Sertore2004CESIUM,lederer2007investigations}; however, several groups have begun to develop mapping technologies to look into the detailed distributions of these key properties over the cathode~\cite{jensen2008application,blaugrund2003measurements,gevorkyan2018effects,karkare2015effects,jensen2014emittance,Riddick2013,zheng2016development,1710.08148,kabra2020mapping,huang2020single}. Several QE mapping technologies have been developed and applied in photoinjectors and non-uniform QE distribution has been observed~\cite{Riddick2013,zheng2016development,1710.08148,kabra2020mapping}. These variations could be caused by localized surface conditions, such as physical and chemical roughness, material defects, and contaminants, etc~\cite{gevorkyan2018effects,Riddick2013}. According to the correlation between QE and thermal emittance~\cite{dowell2009quantum,prat2015measurements,dowell2010cathode,xie2016experimental,moody2018perspectives}, these conditions would also lead to localized variations in thermal emittance. 
	
	To the best of our knowledge, however, there is still a lack in thermal emittance mapping technology for photoinjectors. In fact, measuring the thermal emittance of a fixed cathode area using a single beam remains difficult because all emittance growth factors need to be properly addressed~\cite{miltchev2005measurements,hauri2010intrinsic,prat2015measurements,zheng2016development,zheng2018overestimation}. These factors can be categorized as $\varepsilon _\text{space}$ which denotes the emittance growth from space charge~\cite{limborg2006optimum}, and $\varepsilon_\text{aberration}$ which denotes the growth caused by aberrations, including dipole, quadrupole, and high-order field components inside the radiofrequency (rf) photocathode gun; spherical and chromatic aberrations in the solenoid; and coupled transverse dynamics aberrations~\cite{chae2011emittance,zheng2016development,dowell2018exact,zheng2018overestimation,dowell2018correcting,dowell2016sources,mcdonald1989frontiers,PhysRevAccelBeams.22.072805}. The measurement would be even more challenging and time-consuming when the same procedure is repeated in thermal emittance mapping by scanning a single beam across the cathode with fine steps.
	
	In this paper, we propose and experimentally demonstrate a rapid method to map thermal emittance and QE simultaneously. The basic concept is to use isolated multiple beamlets instead of a single beam in solenoid scan. With proper size and separation of the laser beamlets, the generated electron beamlets are distinguishable in certain current range during the solenoid scan so that their thermal emittance and QE can be measured individually. We believe this method will find broad application in the high-brightness photoinjector community: it can deepen the understanding of the observed non-uniformity, improve simulation accuracy with realistic distribution of cathode properties, and further increase beam brightness by helping to choose the emission site on the cathode.
	
	This paper is organized as follows. Sec.~\ref{section_setup} introduces the beamline layout for analysis and experiment. Sec.~\ref{section2} analyzes the method feasibility with beam dynamics simulation and theoretical derivation. Sec.~\ref{section3} gives more details of the experimental setup. Sec.~\ref{section4} presents the data analysis method and experimental results. Sec.~\ref{section5} briefly discusses the observed correlation between thermal emittance and QE, and studies the current limitations and future improvements of the proposed method. Sec.~\ref{section6} summarizes the current work.
	
	\section{Beamline layout}\label{section_setup}
\begin{figure}[hbtp]
		\centering
		\includegraphics[scale=0.55]{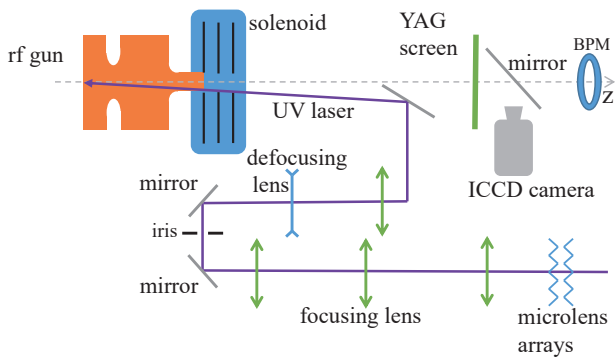}
		\caption{\label{Fig.setup} Beamline setup in simulation and experiment.}
	\end{figure}

	The front end of the drive beamline at Argonne Wakefield Accelerator (AWA) facility~\cite{conde2017research} is used in the beam dynamics analysis and the following experiment, as illustrated in Fig.~\ref{Fig.setup}.
	
	The setup includes an L-band 1.6-cell photocathode gun with a cesium telluride cathode, followed by a solenoid (2.44~m from the cathode) and a screen (2.98~m from the cathode). The incident 248~nm laser has a Gaussian longitudinal distribution with a full width at half maximum (FWHM) pulse length of 1.5~ps. The cathode gradient reaches 62~MV/m in the routine high charge operation~\cite{XueyingPRL2019,shao2020development}. In this study, the cathode gradient is set to 32.5~MV/m to reduce field emission and to improve the signal-to-noise ratio. The laser injection phase is set to $30^\circ$, which is close to the phase of the maximum energy gain under this gradient ($37^\circ$). 
	
	\section{Beam dynamics simulation and theoretical analysis }\label{section2}
	This section first summarizes the basic concept of thermal emittance measurement using the solenoid scan technology with a single beam. Next, it describes the beam dynamics simulation and the theoretical analysis we used to study the feasibility of the proposed thermal emittance mapping method with multiple beamlets. 
	
	In this section, $e$ and $\beta \gamma m c$ denote charge and momentum of electron; $B_0$, $L$, $K \equiv (eB_0)/(2\beta \gamma mc)$, $KL$ denote peak magnetic field, effective length, strength, and Larmor angle of the solenoid, respectively.
	
	\subsection{Solenoid scan with a single beam}
	In the conventional solenoid scan with a single beam, the beam spot size squared taken on a screen downstream of the solenoid varies with the solenoid strength as~\cite{zheng2018overestimation}
	\begin{equation}\label{eq1}
	\begin{aligned}
	\sigma ^2=&(C-L_{d}KS)^{2}\left\langle x_0^2 \right\rangle\\
	&+2(C-L_{d}KS)(S/K+CL_{d})\left\langle {x_0}{x_0}^\prime \right\rangle\\
	&+(S/K+CL_{d})^{2}\left\langle x_{0}^{\prime 2} \right\rangle,\\
	\end{aligned}
	\end{equation}
	where $\left\langle x_0^2 \right\rangle $, $\left\langle {x_0}{x_0}^\prime \right\rangle $ and $\left\langle x_{0}^{\prime 2} \right\rangle $ are the beam moments at the solenoid entrance, $C \equiv \cos (KL)$, $S \equiv \sin (KL)$, and $L_d$ is the drift length between the solenoid exit and the screen. Therefore, the beam moments can be fitted from the beam size when scanning the solenoid strength, and the normalized emittance at the solenoid entrance can be expressed as 
	\begin{equation}\label{eq2}
	{\varepsilon _n} = \beta \gamma \sqrt {\left\langle {x_0^2} \right\rangle \left\langle x_{0}^{\prime 2} \right\rangle  - {{\left\langle {{x_0}{x_0}^\prime } \right\rangle }^2}} ,
	\end{equation}
	
	\subsection{Solenoid scan with multiple beamlets}
	To reduce the amount of time necessary to collect a measurement by moving a single beam for thermal emittance mapping, it is natural to use a pattern beam with isolated multiple beamlets and obtain the thermal emittance distribution in a single-turn measurement. 
	
	\subsubsection{Beam dynamics simulation}
	We first conduct the beam dynamics simulation with the ASTRA code~\cite{floettmann2011astra} to evaluate this concept. The simulation considers three-dimensional field maps with higher-order field components of the photocathode gun and the solenoid~\cite{zheng2018overestimation,PhysRevAccelBeams.22.072805}. The space charge is not considered in the simulation, which is valid for thermal emittance measurements with ultra-low charge.
	
\begin{figure}[hbtp]
		\centering
		\includegraphics[scale=0.5]{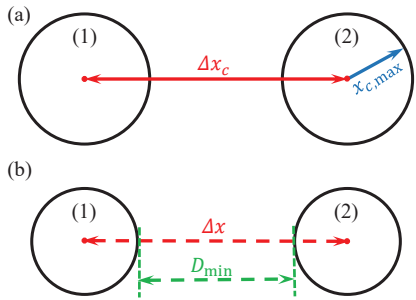}
		\caption{\label{Fig.figure_demonstration} Definition of beamlet geometry parameters on the cathode (a) and on the screen after the solenoid (b). (1) and (2) in the figure indicate the two beamlets.}
	\end{figure}

	For simplicity, we use only two beamlets in the simulation. Each Gaussian-shaped beamlet on the photocathode has a 50~$\mu$m rms spot size and it is cut at 150~$\mu$m radius (3$\sigma$ cut, denoted as $x_{c,\max}$). The first beamlet is set at the center of the cathode, and the second is offset horizontally by 1.812~mm. The center-to-center distance between the beamlets on the cathode is denoted as $\bigtriangleup x_{c}$, as illustrated in Fig.~\ref{Fig.figure_demonstration}(a). The $\varepsilon_{\text{therm},n}$ of both beamlets is set to 1.05~$\mu$m/mm according to a previous experiment~\cite{zheng2018overestimation}. 
	
\begin{figure}[hbtp]
		\centering
		\includegraphics[scale=0.71]{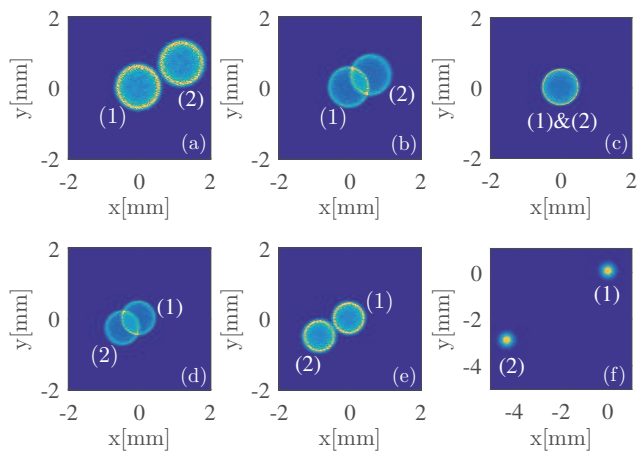}
		\caption{\label{Fig.image_sol_simulation_overlap} The simulated beamlets distribution on the screen as a function of $B_{0}$. In (a-f), $B_{0}$ is set to 0.1859~T, 0.1900~T, 0.1940~T, 0.1970~T, 0.1995~T, and 0.2220~T respectively. (1) and (2) in the figure indicate the two beamlets.}
	\end{figure}

\begin{figure}[hbtp]
		\centering
		\includegraphics[scale=0.64]{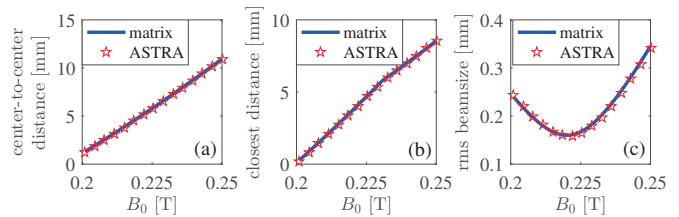}
		\caption{\label{Fig.beamsize_distance_sol} Beam evolution as a function of $B_{0}$ when $B_{0}$ is higher than 0.1995~T. (a) Center-to-center distance. (b) Closest distance between the two beamlets. (c) Rms beam size. Red stars: ASTRA simulation results. Solid blue lines: matrix calculation results. }
	\end{figure}

	The key challenge in solenoid scans with multiple beamlets is to find a suitable range of the solenoid strength for beam moments fitting in which the beamlets are distinguishable. In this range and at the screen position, we define the center-to-center distance between the electron beamlets as $\bigtriangleup x$ and their closest distance as $D_{\min}$, as illustrated in Fig.~\ref{Fig.figure_demonstration}(b). In ASTRA simulation, as the solenoid strength increases from $B_{0}$ of 0.1859~T, the initially separated beamlets [Fig.~\ref{Fig.image_sol_simulation_overlap}(a)] begin to merge [Fig.~\ref{Fig.image_sol_simulation_overlap}(b)], fully overlap [Fig.~\ref{Fig.image_sol_simulation_overlap}(c)], begin to separate [Fig.~\ref{Fig.image_sol_simulation_overlap}(d)], and fully separate again when $B_{0}$ is higher than 0.1995~T [Fig.~\ref{Fig.image_sol_simulation_overlap}(e)]. When $B_{0}$ is higher than 0.1995~T, $\bigtriangleup x$ [Fig.~\ref{Fig.beamsize_distance_sol}(a)] and $D_{\min}$ [Fig.~\ref{Fig.beamsize_distance_sol}(b)] monotonically increase with the solenoid strength. A positive $D_{\min}$ indicates that the beamlets can be well distinguished. The rms size of each individual beamlet $\sqrt{\left\langle {{x^2}} \right\rangle}$ reaches a waist at $B_{0}$ of 0.2220~T [Fig.~\ref{Fig.image_sol_simulation_overlap}(f) and Fig.~\ref{Fig.beamsize_distance_sol}(c)]. 
	
	With the beam spot sizes in the $B_{0}$ range between 0.2 and 0.25~T, the emittance of the two beamlets is fitted to be only 0.2\% (the first beamlet) and 0.3\% (the second beamlet) higher than the thermal value. This negligible emittance growth is caused by $\varepsilon_\text{aberration}$.
	
	\subsubsection{Matrix analysis}
	We then employ a start-to-end matrix calculation of the beam dynamics to analyze the solenoid scan with multiple beamlets and give insight into the ASTRA simulation results.
	
	In a transverse Larmor coordinate (i.e., the axis rotates along the solenoid), the motion of an electron can be expressed as~\cite{wiedemann2015particle}:
	
	\begin{equation}\label{eq100998}
	\left[ {\begin{array}{*{20}{c}}
		x\\
		{{p_x}}
		\end{array}} \right] = \left[ {\begin{array}{*{20}{c}}
		{{R_{11}}}&{{R_{12}}}\\
		{{R_{21}}}&{{R_{22}}}
		\end{array}} \right]\left[ {\begin{array}{*{20}{c}}
		{{x_c}}\\
		{{p_{x_{c}}}}
		\end{array}} \right],
	\end{equation}
	where $x_c$ and $x$ are the beam position on the cathode and the screen, respectively; $p_{x_{c}}=\beta_{x_{c}}\gamma$ and $p_x=\beta_x\gamma$ are the corresponding normalized transverse momenta; and $R_{ij}$ is the transfer matrix element from the cathode to the screen.
	
	Based on Eq.~\ref{eq100998}, the final position $x$ is written as
	\begin{equation}\label{eq765}
	x = {R_{11}}{x_c} + {R_{12}}{p_{x_{c}}}.
	\end{equation} 
	
	For isotropic emission on the cathode with ${{\bar p}_{x_{c}}}=0$ and $\left\langle {{x_c}{p_{x_{c}}}} \right\rangle =0$, the beam position $\bar x$ and the rms beam size square $\left\langle {{x^2}} \right\rangle$ on the screen can be expressed as
	\begin{equation}\label{eq_beamposition}
	\left\{
	\begin{aligned}
	\bar x=&R_{11}\bar {x_c}\\
	\left\langle {{x^2}} \right\rangle=&R_{11}^2\left\langle {x_{c}^2} \right\rangle + R_{12}^2\left\langle {p_{x_{c}}^2} \right\rangle\\
	\end{aligned}
	\right..
	\end{equation}
	
	Therefore, the closest distance between the two beamlets can be expressed as
	\begin{equation}\label{eq_closest_distance}
	D_{\min} = \left| {{R_{11}}\bigtriangleup x_{c}} \right| - 2\left| {{R_{11}}x_{c,\max}} \right| - 2\left| {{R_{12}}{p_{x_{c,\max}}}} \right|,
	\end{equation}
	where $p_{x_{c,\max} }$ is the maximum normalized transverse momentum on the cathode.

\begin{figure}[hbtp]
		\centering
		\includegraphics[scale=0.73]{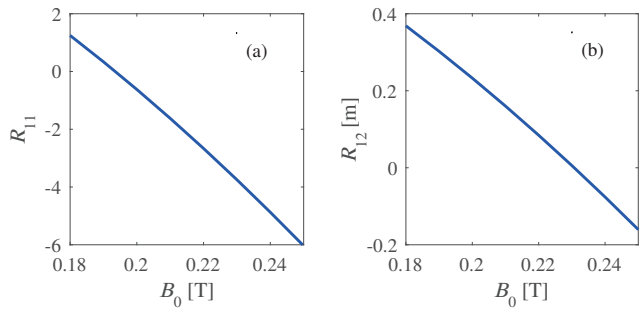}
		\caption{\label{Fig.R11_R12_sol} (a) $R_{11}$ and (b) $R_{12}$ as a function of $B_{0}$.}
	\end{figure}

	We follow the method introduced in Ref.~\cite{gulliford2012new} to calculate the transfer matrix from the cathode to the screen. In this method, the superimposed one-dimensional rf and solenoid fields are considered and the transfer matrix is computed without solving eigenfunction expansions or numerical derivatives. In the calculation, we apply the same beamline parameters as in the ASTRA simulation, except for the three-dimensional field maps. The use of one-dimensional field is valid because $\varepsilon_\text{aberration}$ is negligible based on the previous simulation. The resultant $R_{11}$ and $R_{12}$ are illustrated in Fig.~\ref{Fig.R11_R12_sol}.
	
	The beam parameters are also set to be the same as the ones used in the ASTRA simulation: $\sqrt{\left\langle {x_{c}^2} \right\rangle}$ is 50~$\mu$m, $x_{c,\max}$ is 150~$\mu$m, $\sqrt{\left\langle {p_{x_{c}}^2} \right\rangle}$ is equal to $\varepsilon_{\text{therm},n}$ of 1.05~$\mu$m/mm, $p_{x_{c,\max} }$ is 1.82$\times10^{-3}$ ($\sqrt{3\left\langle {p_{x_{c}}^2} \right\rangle}$ for isotropic emission), and ${{{\bar x}_c}}=0$ and 1.812~mm for the two beamlets ($\bigtriangleup x_{c}$=1.812~mm), respectively.
	
	The calculation has been compared with the ASTRA simulation. The two beamlets fully overlap when $R_{11}=0$. The corresponding $B_{0}$ was found to be 0.194~T based on Fig.~\ref{Fig.R11_R12_sol}(a), which agrees well with the simulation result in Fig.~\ref{Fig.image_sol_simulation_overlap}(c). When the two beamlets are distinguishable ($D_{\min}>0$), their center-to-center distance, the closest distance, and rms beamsize are calculated according to Eq.~(\ref{eq_beamposition}) and Eq.~(\ref{eq_closest_distance}). These equations show good agreement with the ASTRA simulation, as illustrated in Fig.~\ref{Fig.beamsize_distance_sol}.
	
	\section{Experimental setup}\label{section3}
	Fig.~\ref{Fig.setup} illustrates the experimental setup at AWA. At the photocathode gun exit, the electron beams reached 3.3~MeV and were focused by the solenoid onto a retractable YAG screen perpendicular to the beamline. The beam images were reflected by a $45^\circ$ mirror after the YAG screen and captured by a PI-MAX Intensified CCD camera (ICCD) ~\cite{camera}. The shutter width of the camera was set to 100~ns to improve the signal-to-noise ratio. The spatial resolution of the camera was $\sim$60~$\mu$m, measured with a standard USAF target. A calibrated stripline beam position monitor (BPM) downstream was used to measure the charge with a sensitivity of $\sim$40~mV/1~pC \cite{zheng2018overestimation,PhysRevAccelBeams.22.072805}. The minimum detectable charge of the multiple beamlets is therefore 0.05~pC (2~mV).
	
\begin{figure}[hbtp]
		\centering
		\includegraphics[scale=0.8]{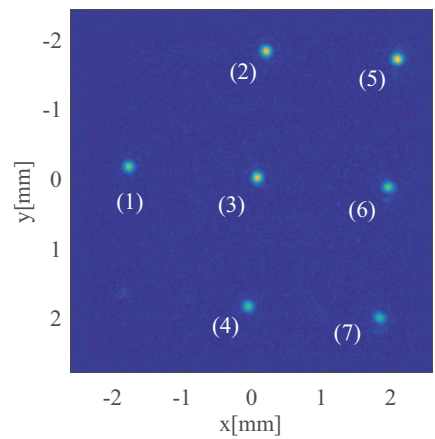}
		\caption{\label{Fig.laser_beam} Laser transverse pattern observed on the virtual cathode. The brightness variation among the laser beamlets was caused by the incident laser, as well as non-ideal conditions of the MLA system. The third beamlet was centered on the cathode.}
	\end{figure}

	The multiple laser beamlets were generated by a microlens-array (MLA) system~\cite{halavanau2017spatial,halavanau2019tailoring}. After passing through a pair of MLAs and three convex lenses, the incident ultraviolet (UV) laser was redistributed to yield a pattern with two-dimensional arrays of beamlets at the iris location. The iris was set to select a few beamlets from the entire pattern. Then the pattern was imaged onto the cathode by a pair of convex and concave lenses. The UV laser energy loss of the MLAs and the lenses was $\sim$90\%. On the virtual cathode (not shown in Fig.~\ref{Fig.setup}), the laser pattern was captured by a UV camera with high spatial resolution of 7.5~$\mu$m/pixel. In this proof-of-principle experiment, the system successfully produced a pattern with seven laser beamlets, as illustrated in Fig.~\ref{Fig.laser_beam}. Each beamlet had a Gaussian-like transverse distribution with an rms spot size of $\sim$50~$\mu$m. The distance between two adjacent beamlets was 1.812~mm. The laser energy of the entire pattern was measured by a UV power meter. The energy of each laser beamlet was calculated according to the beamlet's relative brightness on the UV camera.
	
	\section{Experimental results}\label{section4}
	
	\subsection{Thermal emittance mapping}
\begin{figure}[hbtp]
		\centering
		\includegraphics[scale=0.81]{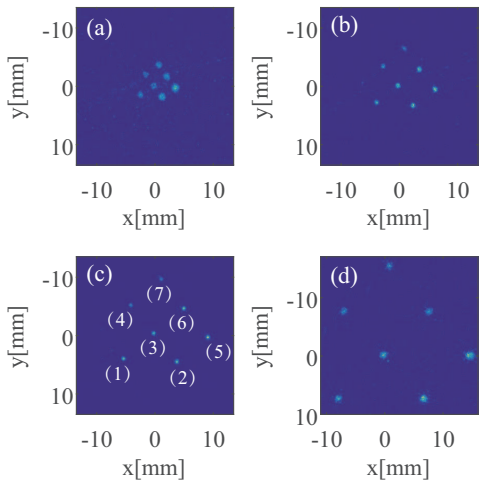}
		\caption{\label{Fig.electron_image} Beam images as a function of solenoid strength. $B_{0}$ in (a)-(d) is 0.2045, 0.2170, 0.2294, and 0.2530~T, respectively. The electron beamlets are marked in (c) following the same laser beamlet index in Fig.~\ref{Fig.laser_beam}.}
	\end{figure}

	The beam images on the screen under different solenoid settings are shown in Fig.~\ref{Fig.electron_image}. The seven beamlets fully overlap when $B_{0}$=0.1940~T, become distinguishable when $B_{0}>$0.2045~T, and reach the waist when $B_{0}$=0.2294~T. These solenoid strengths show reasonable agreement with the ASTRA simulation and the theoretical analysis results.
	
	We calculated the beam size of each electron beamlet using the following procedure. (1) Background field emission images without laser illumination were subtracted to improve the signal-to-noise ratio. (2) Each beamlet was manually selected and its projected distributions in $x$ and $y$ directions were Gaussian fitted to obtain the coarse rms beam sizes, denoted $\sigma_{gx}$ and $\sigma_{gy}$, respectively. (3) For each beamlet, the image within $3\sqrt{\sigma_{gx}\sigma_{gy}}$ of its center was preserved while the rest was set to zero. (4) The cut image was projected in the $x$ and $y$ directions again, and the accurate rms beam sizes were calculated by selecting the central 95\% of the entire area under the projection curve (i.e., 5\% charge-cut of the most outwards part of the beamlet)~\cite{akre2008commissioning}. (5) The rms beam size was calculated as the geometric average of the sizes in the $x$ and $y$ directions: $\sigma=\sqrt{\sigma_{x}\sigma_{y}}$.
	
	Six images were taken under each solenoid strength and we calculated the emittance using the following procedure. (1) For each $B_0$, the rms beam size fluctuated due to machine jitter and its distribution was assumed to be Gaussian. The average value and the standard deviation of the distribution were calculated from the six images as $\bar \sigma$ and $\delta \sigma$, respectively. (2) The average emittance of each beamlet was fitted using $\bar \sigma$ and $B_{0}$ according to Eq.~\ref{eq1}. (3) Another set of beam sizes was generated from the Gaussian distributions and applied to Eq.~\ref{eq1} for emittance fitting. (4) Step (3) was repeated multiple times. The fitted emittance was weighted according to the probability of each set of beam sizes in the Gaussian distributions, from which the standard deviation of emittance can be calculated. It should be noted that the error bar in fitting is negligible when compared to the standard deviation caused by machine jitter in our measurement.
	
	\begin{figure}[hbtp]
		\centering
		\includegraphics[scale=0.8]{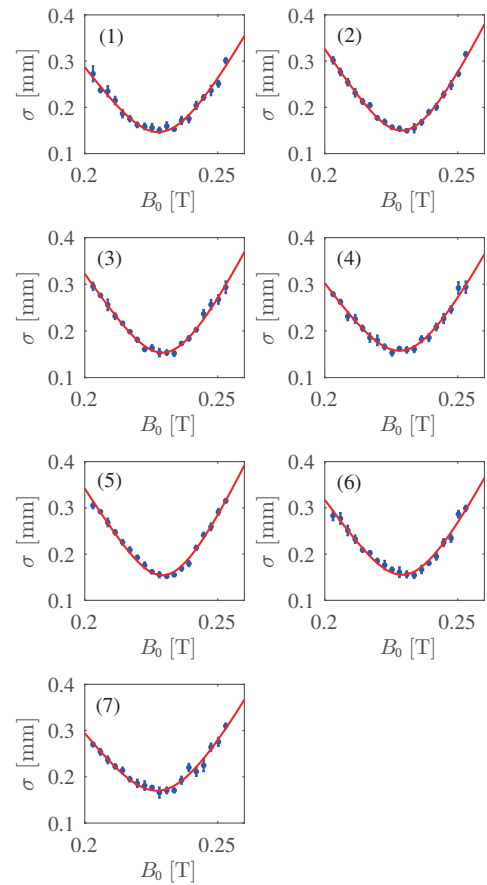}
		\caption{\label{Fig.fitting_curve} The rms beam sizes of the seven beamlets (marked by number) as a function of $B_{0}$. Blue dots: experimental data. The error bar at each $B_{0}$ denotes the standard deviation of the measured beam size. Red lines: fitting with $\bar \sigma$ according to Eq.~\ref{eq1}.}
	\end{figure}
	
	The influence of $\varepsilon_\text{aberration}$ on the thermal emittance measurement has been studied in ASTRA using the same settings as in Sec.~\ref{section2}.B.1. The simulation results indicate the influence is negligible: the emittance growth is 0.2\% for beamlet 3 (at the cathode center); 0.3\% for beamlets 1, 2, 4, and 6 (1.812~mm off-axis); and 0.4\% for beamlets 5 and 7 (2.563~mm off-axis).
	
\begin{figure}[hbtp]
		\centering
		\includegraphics[scale=0.7]{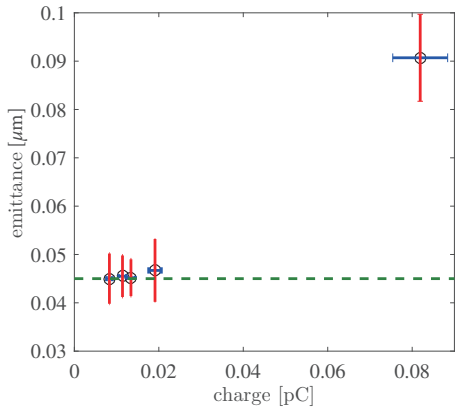}
		\caption{\label{Fig.emt_charge} Measured emittance of beamlet 1 as a function of beamlet charge. Black circles with error bars: experimental data. Dashed green line: converged emittance.}
	\end{figure}

	The influence of $\varepsilon_\text{space}$ was experimentally minimized by reducing the total charge of the pattern beam. The total incident laser energy was controlled by neutral density filters (NDFs) before the MLA system. The charge of each beamlet can be calculated from the total charge measured by the calibrated BPM and the relative brightness of the beamlet over the entire pattern on the YAG screen. For example, the measured emittance of beamlet 1 as a function of its charge is shown in Fig.~\ref{Fig.emt_charge}. The measured emittance converged when the beamlet charge was lower than 0.02~pC, which indicates that $\varepsilon _\text{space}$ is negligible and $\varepsilon _{n} \approx \varepsilon_\text{therm}$ below this charge level.
	
	The measured thermal emittance of the seven beamlets is summarized in Table~\ref{t12}. In this table, the spot size of each laser beamlet is calculated following the same steps (steps 2-5) as the electron beamlets. Its average value and standard deviation are calculated using 20 images taken by the UV camera. The resultant thermal emittance varies from 0.934 to 1.142~$\mu$m/mm. Their average value is close to the previous experimental result of 1.05~$\mu$m/mm using a large single beam with 3~mm diameter~\cite{zheng2018overestimation}. We notice that these values are higher than some of the previously reported results of thermal emittance of cesium telluride cathodes~\cite{Sertore2004CESIUM,Eduard2015Measurements,FilippettoCesium}. This may be caused by various factors, such as cathode preparation procedures, surface conditions, and operation conditions.
	
	\begin{table*}[hbtp]
		\caption{\label{t12} Experimental results of rms spot size of the laser beamlet $\sigma_\text{laser}$, thermal emittance $\varepsilon _\text{therm}$, $\varepsilon _{\text{therm},n}$, and QE of the seven beamlets. The error bar denotes the standard deviation of the measurement.}
		\renewcommand\tabcolsep{16pt}
		\renewcommand\arraystretch{1.3}
		\begin{tabular}{ccccc}
			\toprule[1.5pt]
			\multicolumn{1}{c}{Beamlet No.} & \multicolumn{1}{c}{$\sigma_\text{laser}$ {(}$\mu$m{)}} & \multicolumn{1}{c}{$\varepsilon _\text{therm}$ {(}$\mu$m{)}} & \multicolumn{1}{c}{$\varepsilon _{\text{therm},n}$ {(}$\mu$m/mm{)}} &\multicolumn{1}{c}{QE(\%)}\\ 
			\midrule[1pt]
			1                       & 48.4 $\pm$ 0.5                                     & 0.0452 $\pm$ 0.0038                        & 0.934 $\pm$ 0.079        & 4.59 $\pm$ 0.66 \\
			2                       & 49.8 $\pm$ 0.4                                     & 0.0518 $\pm$ 0.0031                        & 1.040 $\pm$ 0.063        & 7.17 $\pm$ 0.80 \\
			3                       & 50.2 $\pm$ 0.4                                     & 0.0515 $\pm$ 0.0034                        & 1.026 $\pm$ 0.068        & 6.15 $\pm$ 0.57 \\
			4                       & 49.7 $\pm$ 0.7                                     & 0.0500 $\pm$ 0.0041                        & 1.006 $\pm$ 0.084        & 5.60 $\pm$ 0.79 \\
			5                       & 48.7 $\pm$ 0.5                                     & 0.0556 $\pm$ 0.0026                        & 1.142 $\pm$ 0.055        & 8.66 $\pm$ 0.63 \\
			6                       & 49.9 $\pm$ 0.5                                     & 0.0510 $\pm$ 0.0042                        & 1.022 $\pm$ 0.085        & 5.64 $\pm$ 0.79 \\
			7                       & 51.8 $\pm$ 0.9                                     & 0.0520 $\pm$ 0.0033                        & 1.004 $\pm$ 0.066        & 5.47 $\pm$ 0.55 \\  \bottomrule[1.5pt]
		\end{tabular}
	\end{table*}
	
	\subsection{Quantum efficiency (QE) mapping}
\begin{figure}[hbtp]
		\centering
		\includegraphics[scale=0.7]{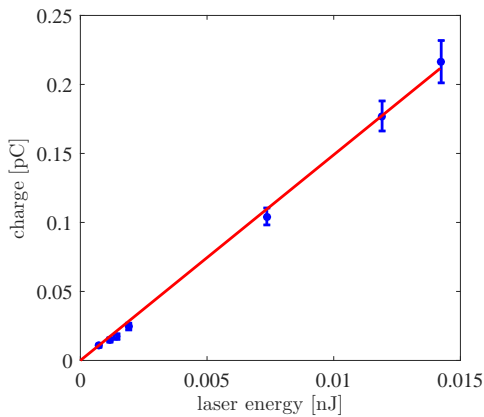}
		\caption{\label{Fig.charge_laser_energy} Measured charge as a function of laser energy for beamlet 1. Blue dots: experimental data. The error bar denotes the standard deviation of the measured charge. Red line: linear fitting.}
	\end{figure}

	The QE mapping is straightforward with isolated multiple beamlets, because the laser energy and the charge of each beamlet can be derived based on experimental
	measurement. For example, Fig.~\ref{Fig.charge_laser_energy} shows the charge of beamlet 1 when gradually reducing the laser energy with NDFs. The linearity between the charge and the laser energy confirms single-photon emission, and its slope is used to calculate the QE. The QE of the seven beamlets is also summarized in Table.~\ref{t12}.
	
	\section{Discussion}\label{section5}
	\subsection{Dependence of thermal emittance on QE}
\begin{figure}[hbtp]
		\centering
		\includegraphics[scale=0.7]{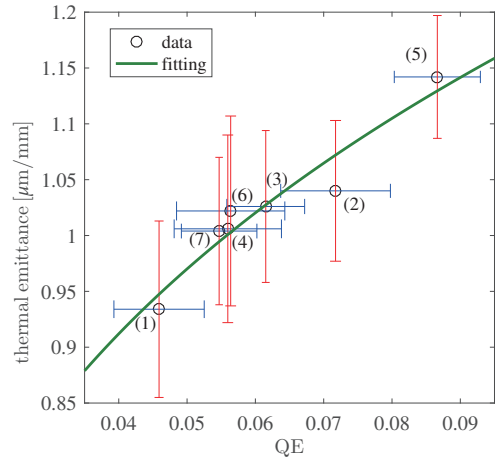}
		\caption{\label{Fig.emt_QE} Dependence of thermal emittance on QE. Black circles with error bars: experiment data of the seven beamlets (marked by number). Green line: least-square fitting.}
	\end{figure}

	Based on Table~\ref{t12}, we can see that beamlets with higher QE usually have higher thermal emittance. This trend is illustrated in Fig.~\ref{Fig.emt_QE}.
	
	In a metallic photocathode, the thermal emittance and QE depend on electron excess energy $h\nu-\phi _\text{eff}$. This can be described by Dowell's three-step model~\cite{dowell2009quantum} as follows:
	\begin{equation}\label{eq_correlation}
	\left\{
	\begin{aligned}
	\varepsilon _{\text{therm},n} \propto & \sqrt {h\nu  - {\phi _\text{eff}}} \\
	\text{QE} \propto & (h\nu  - {\phi _\text{eff}})^2 \\
	\end{aligned}
	\right.,
	\end{equation}
	where $h\nu$ is the incident photon energy and $\phi _\text{eff}$ is the effective work function.
	
	In a semiconductor photocathode, the first equation holds valid by assuming that most photoelectrons emit from the valence band~\cite{prat2015measurements,dowell2010cathode}. However, the second one is still controversial due to the complicated emission mechanism~\cite{xie2016experimental,moody2018perspectives}. Ref.~\cite{moody2018perspectives} suggested a modified version:
	\begin{equation}\label{eq13}
	\text{QE} \propto (h\nu  - {\phi _\text{eff}})^p,
	\end{equation}
	where $p$ is a constant.
	
	From the experimental data, the dependence of thermal emittance on QE was least-square fitted as $\varepsilon _{\text{therm},n}  = 2.223\times\sqrt[3.61]{\text{QE}}$, which resulted in $p=1.805$. This number is within the large range of $p$ from 1.3 to 4.6 extracted from previous studies~\cite{Heikophdthesis,gaowei2019codeposition,wisniewski2013kelvin}. The large variation of $p$ may be caused by preparation procedures, surface conditions, operation conditions, and other factors. The physical understanding of the constant $p$ requires more fundamental research in the future.
	
	\subsection{Mapping limitation and future improvement}
	We measured the thermal emittance and QE of seven isolated spots near the cathode center in this proof-of-principle experiment. In future studies, the mapping method could be improved with larger area, higher resolution, and higher density.
	
\begin{figure}[hbtp]
		\centering
		\includegraphics[scale=0.75]{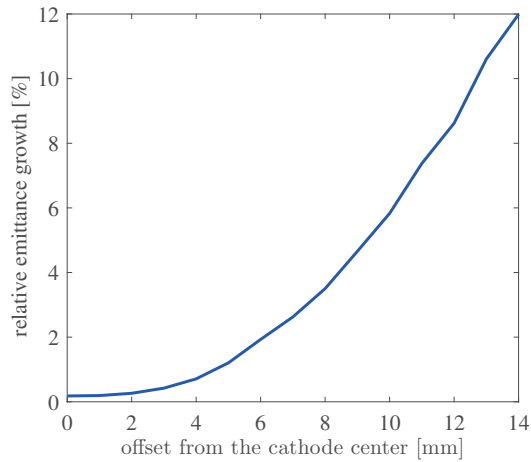}
		\caption{\label{Fig.emittance_growth_offset} ($\varepsilon _{n}-\varepsilon _{\text{therm}})/\varepsilon _{\text{therm}}$ as a function of the beamlet offset on the cathode.}
	\end{figure}

	The mapping area is defined as the maximum boundary that the beamlets can cover on the cathode. When the space charge effect is negligible, the mapping area will be limited by the emittance growth from the aberration terms. We use ASTRA simulation with a single off-axis beamlet to evaluate the growth. In the simulation, the beamline parameters and the size/emittance of the beamlet are kept the same as the ones in Sec.~\ref{section2}.B.1. A quad corrector, containing a pair of normal and skew quadrupoles, has been added to correct the coupled transverse dynamics aberration. More details of the quad corrector can be found in Ref.~\cite{zheng2018overestimation} and Ref.~\cite{PhysRevAccelBeams.22.072805}. The simulation results show monotonic emittance growth when increasing the beam offset, as illustrated in Fig.~\ref{Fig.emittance_growth_offset}. The emittance growth reaches 10\% when the beam offset is 12.5~mm.
	
	The mapping resolution is defined as the smallest area that can be measured by a beamlet. It can be improved by using smaller laser spots in general, but will be eventually limited by the system resolution in a given experiment, such as the UV camera that determines the laser beamlet spot size and the YAG screen that determines the electron beamlet spot size.
	
	The mapping density is defined as the number of sampling beamlets per unit area. Unlike some QE mapping methods, in which a single beam scans across the cathode continuously, the initial beamlets in the proposed method need to be separated by a certain distance when placed on the cathode. The minimum separation, or the highest density, is physically limited by the beamlet overlapping issue.
	
\begin{figure}[hbtp]
		\centering
		\includegraphics[scale=0.7]{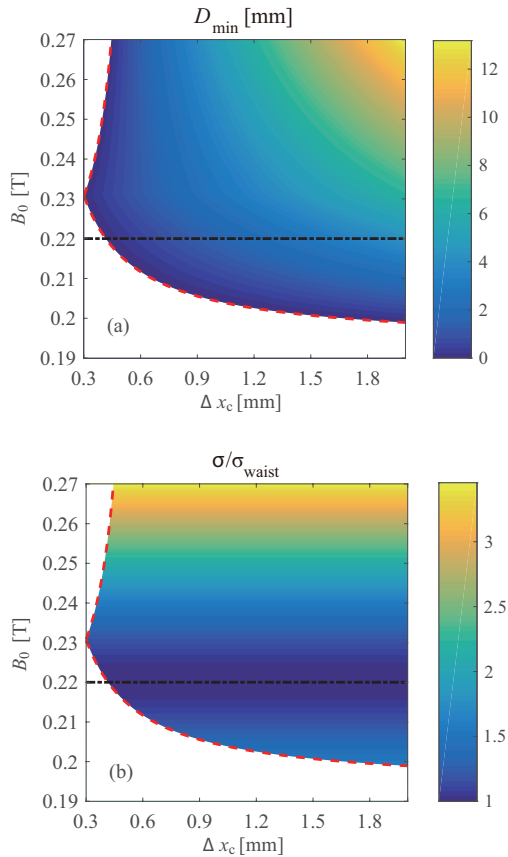}
		\caption{\label{Fig.Dmin_B0_xc} Dependence of (a) $D_{\min}$ and (b) $\sigma /\sigma _{\text{waist}}$ on $\bigtriangleup x_{c}$ and $B_{0}$. The white areas indicate the beamlets cannot be distinguished. Black dotted line: solenoid strength when beam size reaches a waist.}
	\end{figure}

	Based on Eq.~\ref{eq_closest_distance}, the dependence of the closest distance between two adjacent electron beamlets $D_{\min}$ on the laser beamlets separation $\bigtriangleup x_{c}$ and the solenoid strength $B_{0}$ can be calculated, as illustrated in Figure.~\ref{Fig.Dmin_B0_xc}(a). The beamline and beam parameters, other than the initial separation, remain the same as those in Sec.~\ref{section2}.B.1. The minimum initial distance in the calculation is set to 300~$\mu$m, which is twice the beamlet radius. When the initial separation is larger, there will always be a $B_{0}$ range in which the beamlets are distinguishable ($D_{\min}>0$). In principle, the beam moments can be fitted out in this range and the emittance can be calculated accordingly. In practice, however, it is desirable to include both sides around the beam waist in fitting, in order to obtain accurate results~\cite{bazarov2011thermal}, as illustrated in Fig.~\ref{Fig.beamsize_distance_sol}(c) and Fig.~\ref{Fig.fitting_curve}. In Fig.~\ref{Fig.Dmin_B0_xc}(a), when $\bigtriangleup x_{c}$ is larger than 415~$\mu$m, both sides around the beam waist have a distinguishable range. 
	
\begin{figure}[hbtp]
		\centering
		\includegraphics[scale=0.65]{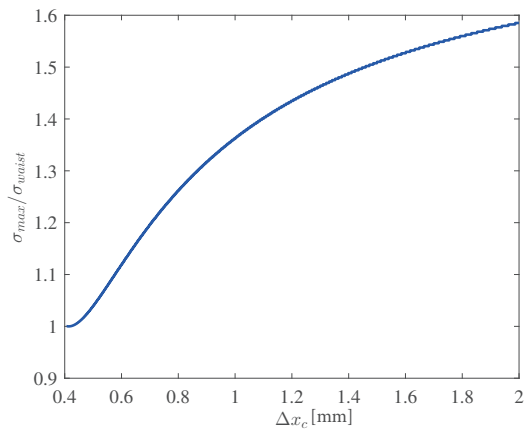}
		\caption{\label{Fig.xc_completeness} Maximum $\sigma /\sigma _{\text{waist}}$ at the low $B_{0}$ end of the distinguishable range as a function of $\bigtriangleup x_{c}$.}
	\end{figure}

	Even when both sides are available, there are still several different criteria to select the $B_{0}$ range to improve the fitting accuracy~\cite{miltchev2006investigations,houjunphdthesis,schmeisser2018situ}. For example, Ref.~\cite{houjunphdthesis} selects the range in such a way that the maximum beam sizes at both sides are twice of the waist. From Fig.~\ref{Fig.Dmin_B0_xc}(b), we learn that the maximum beam size for certain $\bigtriangleup x_{c}$ is limited by the low $B_{0}$ end. In general, larger $\bigtriangleup x_{c}$ leads to larger beam size at this end, as illustrated in Fig.~\ref{Fig.xc_completeness}. Further discussion of the range selecting criteria is beyond the scope of the current study. It should be noted that, in addition to the selecting criteria, the minimum separation also depends on the initial laser spot size and the thermal emittance.
	
	\section{Summary}\label{section6}
	In summary, we introduce a rapid thermal emittance and QE mapping method that uses multiple beamlets in solenoid scan. Its feasibility is supported by beam dynamics simulations and theoretical analysis. In a proof-of-principle experiment using an L-band rf photoinjector with a cesium telluride cathode, seven beamlets with 50~$\mu$m rms beam size and 1.812~mm separation were generated and their thermal emittance (varying from 0.93 to 1.14~$\mu$m/mm) and QE (varying from 4.6\% to 8.7\%) were successfully measured. The range, resolution, and the density of the proposed method can be improved by using smaller laser beamlets with denser separation in a larger area, which could be achieved by optimizing the MLA system. The ultimate performance will be limited by the emittance growth from the aberration terms, the system resolution, and the beamlet overlapping issue.

	
	\begin{acknowledgments}
		The work at AWA is funded through the U.S. Department of Energy (DOE) Office of Science under Contract No. DE-AC02-06CH11357. The work is also supported by the National Natural Science Foundation of China (NSFC) No. 11435015 and No. 11375097.
	\end{acknowledgments}
	
	\bibliography{apstemplate.bib}
	
\end{document}